\documentclass{Interspeech2024}

\usepackage{graphicx}
\usepackage{multirow}
\usepackage{multicol}
\usepackage{bm}
\usepackage{balance}
\usepackage{bbding}

\interspeechcameraready 

\let\svthefootnote\thefootnote
\newcommand\blankfootnote[1]{%
  \let\thefootnote\relax\footnotetext{#1}%
  \let\thefootnote\svthefootnote%
}

\title{Retrieval Augmented Generation in Prompt-based Text-to-Speech Synthesis with Context-Aware Contrastive Language-Audio Pretraining}

\name[affiliation={\dagger}]{Jinlong}{Xue}
\name[affiliation={\dagger}]{Yayue}{Deng}
\name[affiliation={}]{Yingming}{Gao}
\name[affiliation={*}]{Ya}{Li}

\email{\{jinlong\_xue, yayue.deng, yingming.gao, yli01\}@bupt.edu.cn}

\address{
  Beijing University of Posts and Telecommunications, Beijing, China
  }


\begin{document}

\maketitle

\blankfootnote{$\dagger$ Equal Contribution. * Corresponding author.}

\vspace{-0.2cm}

\begin{abstract}

Recent prompt-based text-to-speech (TTS) models can clone an unseen speaker using only a short speech prompt. They leverage a strong in-context ability to mimic the speech prompts, including speaker style, prosody, and emotion. Therefore, the selection of a speech prompt greatly influences the generated speech, akin to the importance of a prompt in large language models (LLMs). However, current prompt-based TTS models choose the speech prompt manually or simply at random. Hence, in this paper, we adapt retrieval augmented generation (RAG) from LLMs to prompt-based TTS. Unlike traditional RAG methods, we additionally consider contextual information during the retrieval process and present a Context-Aware Contrastive Language-Audio Pre-training (CA-CLAP) model to extract context-aware, style-related features. The objective and subjective evaluations demonstrate that our proposed RAG method outperforms baselines, and our CA-CLAP achieves better results than text-only retrieval methods.

\end{abstract}

\vspace{-0.2cm}
\section{Introduction}


Text-to-speech (TTS) synthesis aims to generate natural speech from text and has seen tremendous improvements due to the adoption of deep learning methods. Recently, the integration of Large Language Models (LLMs) with TTS synthesis technology has emerged as a new trend, garnering widespread attention. LLMs, through in-context learning (ICL), have shown significant advancements in learning from minimal prompts. This breakthrough, coupled with the use of neural audio codecs~\cite{soundstream,encodec,hubert} that convert continuous audio features into discrete tokens, has greatly propelled recent speech synthesis frameworks~\cite{megatts,megatts2}, such as VALL-E~\cite{valle}, AudioLM~\cite{audiolm}, NaturalSpeech2~\cite{naturespeech2}, and SPEAR-TTS~\cite{speartts}. These systems can generate high-quality, personalized speech from just a few seconds of unseen audio used as a speech prompt.


VALL-E~\cite{valle}, a pioneering TTS framework, adopts RVQ-based audio codec Encodec~\cite{encodec} and utilizes a language model as a prompt-based language modeling task. It can generate acoustic tokens based on the input of a only 3-second voice recording. AudioLM~\cite{audiolm} uses a hierarchical sequence-to-sequence approach and adopts w2v-BERT~\cite{w2vbert} and SoundStream~\cite{soundstream} to extract semantic and acoustic tokens respectively. Therefore the speech prompt is used in both stages and extracted with different representations. SPEAR-TTS~\cite{speartts} has the same structure except for replacing the first stage with an encoder-decoder scheme. Compared with traditional TTS systems like FastSpeech2~\cite{fastspeech2} and Tacotron2~\cite{tacotron}, these recent models show great voice cloning ability by providing only a 3-second speech prompt and have natural prosody comparable with human speakers. This huge success can be attributed to in-context learning provided by GPT-like architecture, and adoption of audio codecs which enable TTS models to utilize vast, diverse, and noisy data instead of only recorded data. However, the generation in a GPT-like manner is highly dependent on previously predicted tokens. This means that the speech prompt has a substantial impact on the subsequent generation process, significantly influencing the generated speech and affecting aspects such as speaker timbre, prosody, and speaking style.




\begin{figure*}[th!]
  \centering
  \includegraphics[width=.95\linewidth]{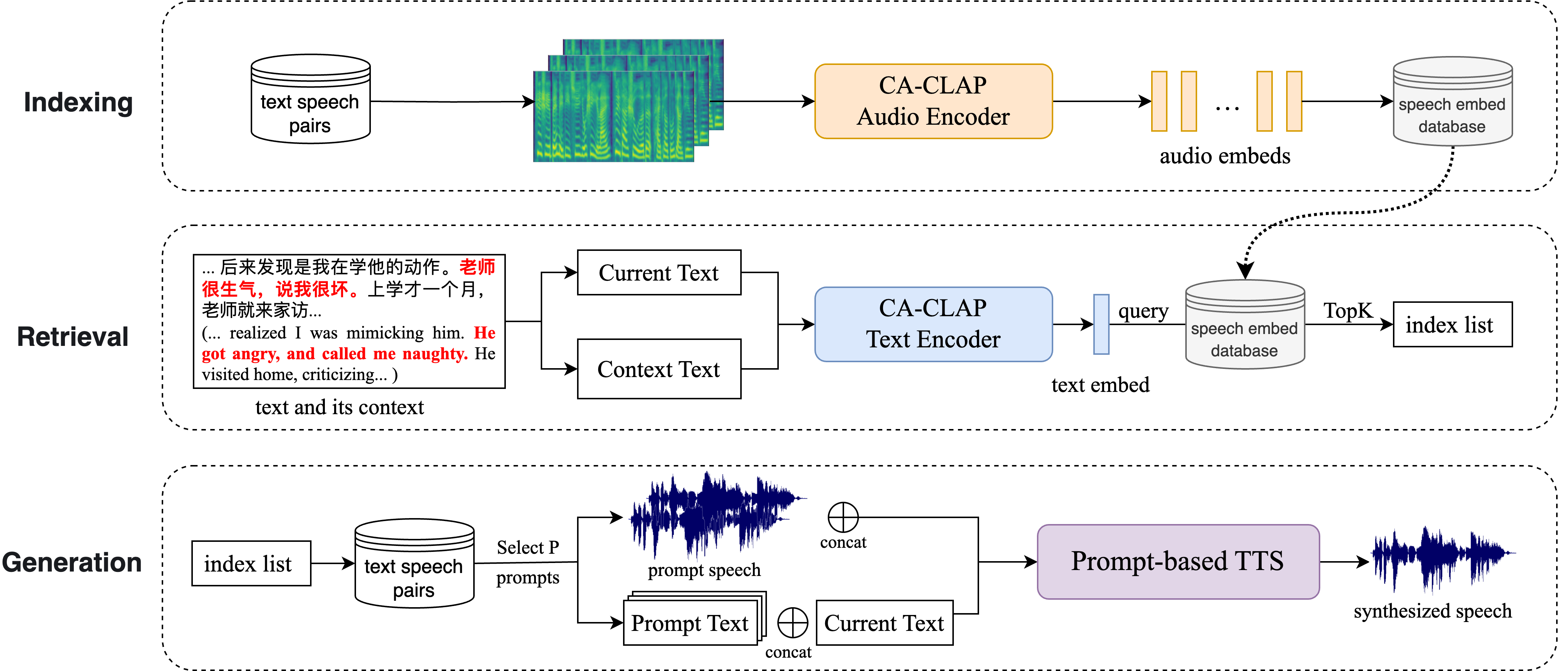}
  \caption{An overview of our RAG-enhanced prompt-based TTS}
  \label{fig:overview}
  \vspace{-0.4cm}
\end{figure*}


Hence, the selection of speech prompts is critically important, akin to the significance of prompts in the LLM domain, where the quality and clarity of prompts significantly influence the outcomes~\cite{cot,zerocot}. However, existing methods often randomly select speech prompts from the target speaker, resulting in a choice that is frequently inadequate for guiding the zero-shot TTS system to mimic the desired speaking style and target timbre effectively. The choice of speech prompt should vary given different texts. Furthermore, in TTS scenarios that incorporate context information, such as audiobook TTS~\cite{xue2022paratts,chen2022unsupervised,chen2023stylespeech} and conversational TTS~\cite{m2ctts,deng2023cmcu,deng2023concss}, the choice of a speech prompt should also take contextual information into account. 


To address this challenge, the given audio prompt should coherent the style information with current text and context information. In LLM area, RAG methods~\cite{rag4ki,ragmp} are recognized as a significant enhancement across a variety of tasks. Since LLMs cannot accurately memorize every piece of knowledge but they have strong in-context learning abilities, RAG methods find the most relevant information from external databases and use them as prompts. Retrieval augments the LLM’s ability to generate accurate, grounded responses, especially for queries demanding specialized domain knowledge. 

Motivated by this insight, we adapt the RAG concept to the speech domain to tackle the challenge of selecting appropriate speech prompts. To this end, we introduce a novel framework that combines context-aware retrieval-augmented generation with a prompt-based TTS system. Furthermore, unlike traditional RAG methods that rely solely on textual data, our approach incorporates acoustic inputs during retrieval. This is because the acoustic modality offers richer information, including emotion and speaking style, enhancing the overall quality and relevance of the retrieved content. Specifically, our proposed framework incorporates an innovative Context-Aware Contrastive Language-Audio Pre-training (CA-CLAP) model which is designed to extract context-aware, style-related textual features (STFs) under audio supervision. It employs an audio encoder for extracting style embeddings from speech and a text encoder for deriving STFs from both the text and its context. Additionally, we enhance context integration by implementing cross-attention mechanisms between textual and contextual features. Overall, our paper makes the following contributions: 1) We propose a RAG-enhanced prompt-based TTS framework to enhance audio prompt specialized selection. 2) We design a CA-CLAP model to extract textual and acoustic representations for retrieval. 3) We conduct extensive subjective and objective experiments and find that our proposed methods outperform baselines and our introduced CA-CLAP has better results than text-only embedding methods. Audio samples are available on the project page\footnote{https://happylittlecat2333.github.io/interspeech2024-RAG}.

\vspace{-0.3cm}
\section{Methodology}

The details of our proposed RAG-enhanced method, along with the introduction of our CA-CLAP and the prompt-based TTS are described in the below sections.

\vspace{-0.2cm}
\subsection{RAG-enhanced Prompt-based TTS}

As shown in Fig.~\ref{fig:overview}, our proposed method includes three components: indexing, retrieval, and generation. The indexing process is a crucial initial step that transfers all the possible audio prompts into style-related representations via the audio encoder of pretrained context-aware contrastive language-audio model (CA-CLAP) and subsequently constructs a speech embedding database. It serves as key similarity comparison during the retrieval phase. Then, in the retrieval process, the current text and context are encoded into the style-related text features (STFs) via the pretrained CA-CLAP text encoder. The generated context-aware text representations are served as query to compute the similarity scores with the vectorized audio prompts within the indexed speech embedding database. Then, the model prioritizes and retrieves the top K audio prompts that demonstrate the greatest similarity. Finally, in the generation process, we use the first P prompts and concatenate them as final audio prompts, to guide the pretrained prompt-based TTS system to generate suitable speech.


\subsection{Context-Aware Contrastive Language-audio Pretraining (CA-CLAP)}

\begin{figure*}[h!]
  \centering
  \includegraphics[width=.95\linewidth]{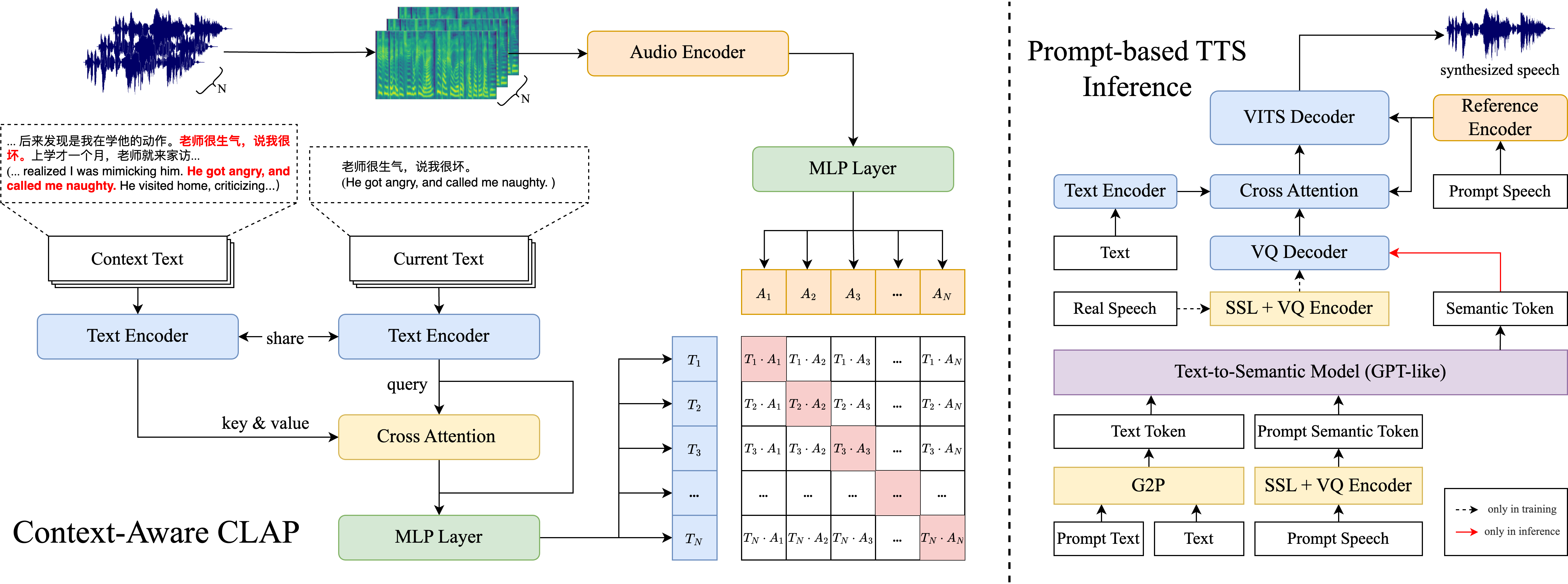}
  \caption{Left: An Overview of Context-Aware CLAP. Right: An illustration of the inference process of Prompt-based TTS}
  \label{fig:CA-CLAP}
  \vspace{-0.3cm}
\end{figure*}

In the indexing process of the RAG-enhanced prompt-based TTS structure, textual modality and acoustic modality inputs (text and audio) need to be embedded into a shared feature space to calculate the cosine similarity. Hence, inspired by~\cite{clap}, we consider constructing a multi-modal feature extractor which can extract style-related embedding from both textual and acoustic inputs. We implement two encoders to separately process audio data and text data. Moreover, to fully utilize the additional contextual information, we enhance the current linguistic feature with contextual information via a cross-attention mechanism. The context definition and computation formula are as: 

\vspace{-0.3cm}
\begin{equation}
    U_{con}=Concat(U_{i-l}, U_{i-l+1}, ... , U_{i}, ..., U_{i+l-1}, U_{i+l})
\end{equation}
\begin{equation}
    Q = W^Q E_{cur}, \quad
    K = W^K E_{con}, \quad
    V = W^V E_{con} 
\end{equation}
\begin{equation}
    CrossAtt(E_{cur},E_{con}) = Softmax (\frac{Q K^{T}}{\sqrt{d}}) \cdot V
\end{equation}

\noindent where $l$ is the context length, $U_i$ is current text and $U_{con}$ is context text. $E_{cur}$ and $E_{con}$ denote the SFTs from current text and context text respectively. $W^K$, $W^Q$ and $W^V$ represent the weight matrix of attention key, query and value respectively. Our implementation adopts a shared text encoder for both current text and context text. The cross-attention mechanism is applied to the text encoder outputs with style-related text features (STFs) from current text $E_{cur}$ as query $Q$ and STFs from context $E_{con}$ as key $K$ and value $V$. 

In short, the proposed CA-CLAP model serves as an encoding model during both indexing and retrieval phases of RAG, transferring multi-modal inputs into a shared feature space based on context-aware contrastive learning. Therefore context-aware SFTs can retrieve the relevant audios as speech prompts.

During training, as illustrated in the left part of Fig.~\ref{fig:CA-CLAP}, given N (speech, text) pairs as input, CA-CLAP computes an $N \times N$ matrix M. The value at the $i$-th row and $j$-th column represents the cosine similarity between the text embedding $T_i$ of the $i$-th text, obtained by the CA-CLAP text encoder, and the audio embedding $A_j$ of the $j$-th speech, obtained by the CA-CLAP audio encoder. The text and audio encoders strive to maximize the cosine similarity for the $N$ correct pairs in the batch and minimize it for the $N^2-N$ incorrect pairings.

The model is trained with the contrastive learning paradigm between the audio and text embeddings in pairs, following the same loss function as in~\cite{clap}:

\begin{footnotesize}
\begin{align}    
L = \frac{1}{2N} \sum_{i=1}^{N} ( \log \frac{\exp (A_{i} \cdot T_{i} / \tau)}{\sum_{j=1}^{N} \exp (A_{i} \cdot T_{j} / \tau)} \nonumber \\
 + \log \frac{\exp (T_{i} \cdot A_{i} / \tau)}{\sum_{j=1}^{N} \exp(T_{i} \cdot A_{j} / \tau)} )
\label{equ:1}
\end{align}
\end{footnotesize}

\noindent where $\tau$ is a learnable temperature parameter for scaling the loss. Two logarithmic terms consider either audio-to-text logits or text-to-audio logits. $N$ is used as the batch size. The relevance of text-audio pair embeddings is scored by the cosine distance calculation.



\vspace{-0.2cm}
\subsection{Prompt-based Text-to-Speech}

The backbone of the prompt-based speech synthesis model we employ is GPT-SoVITS, as shown in the right part of Fig.~\ref{fig:CA-CLAP}. It uses discrete semantic tokens as an intermediate feature between the VITS~\cite{vits} decoder and the text-to-semantic model. The model leverages a self-supervised learning (SSL) model HuBERT~\cite{hubert} to extract discrete semantic tokens. Additionally, it incorporates a reference encoder from TransferTTS~\cite{transferTTS} to extract speaker embeddings.

The training process is divided into two stages. In the first stage, the VITS decoder and the vector quantization (VQ) model are jointly trained with VITS loss and VQ commitment loss. In the second stage, utilizing the well-trained VQ model from the first stage and the pretrained HuBERT model, the text-to-semantic model is trained in a GPT-like manner to predict the next semantic token.

\vspace{-0.2cm}
\section{Experiments}

\subsection{Training Setup}

To train our proposed CA-CLAP model, we collect 3254 Chinese audiobooks from Internet, including 616 hours. We first separate the background music and split the whole speech into utterances, then we use Paraformer\footnote{https://github.com/alibaba-damo-academy/FunASR} to transcribe the audio. We split our collected dataset with 100 audiobooks including 9K text-audio pairs for test and 10 audiobooks for validation, and other 3144 audiobooks including 285K text-audio pairs for training. We use the previous and following 5 sentences of the current text as the corresponding contextual content. We use RoBERTa\footnote{https://huggingface.co/hfl/chinese-roberta-wwm-ext} as text encoder and HTSAT~\cite{chen2022htsat} as audio encoder. We train our CA-CLAP model for 10 epochs on one NVIDIA A6000 GPU with a batch size of 120. 

To effectively evaluate our proposed RAG method, we additionally collect 48 audiobooks that have at least 500 utterances to ensure there are sufficient samples to retrieve. We use the last 100 utterances for test and use the other utterances in the same book for retrieval. Therefore, we have 4800 utterances for evaluation. We adopt the well-performed prompt-based TTS model GPT-SoVITS\footnote{https://github.com/RVC-Boss/GPT-SoVITS} as our TTS backbone and use retrieved prompt text-audio pairs to provide prosody and speaker timbre.


\begin{table*}[t!]
\centering
\caption{Objective and subjective experiments comparing zero-shot TTS performance across various retrieval methods. `Text' and `Audio' indicate the modalities utilized for retrieval. We evaluate the 95\% confidence intervals for NMOS and SMOS.}
\scalebox{.85}{
\begin{tabular}{l|c|c|c|c|c|c|c|c}
\hline \hline
 & \multicolumn{2}{c|}{\textbf{Retrieval}} &  \multicolumn{4}{c|}{\textbf{Objective}} & \multicolumn{2}{c}{\textbf{Subjective}} \\
\hline
\textbf{Methods} & Text & Audio & Energy$\downarrow$ & F0$\downarrow$ & MCD$\downarrow$ & SECS$\uparrow$ & NMOS$\uparrow$  & SMOS$\uparrow$ \\
\hline
Groundtruth & - & - & -  & - & - & -  & 4.408 $\pm$ 0.089 & -    \\
Self        & - & - & 15.144 & 46.785 & 5.664 & 0.806 & 4.110 $\pm$ 0.058 & 4.002 $\pm$ 0.072 \\
\hline
Random      & $\times$ & $\times$  & 17.877 & 59.007 & 6.689 & 0.724 & 3.624 $\pm$ 0.082 & 3.314 $\pm$ 0.077 \\
MiniLM      & \checkmark & $\times$ & 17.612 & 58.271 & 6.583 & 0.730 & 3.681 $\pm$ 0.119 & 3.525 $\pm$ 0.082 \\
CoROM       & \checkmark & $\times$ & 17.605 & 57.930 & 6.577 & 0.731 & 3.766 $\pm$ 0.068 & 3.455 $\pm$ 0.109 \\
\hline
Proposed    & \checkmark & \checkmark & \textbf{17.333} & \textbf{57.678} & \textbf{6.507} & \textbf{0.734} & \textbf{3.922} $\pm$ \textbf{0.068} & \textbf{3.778} $\pm$ \textbf{0.075} \\
\hline \hline
\end{tabular}
}
\vspace{-0.4cm}
\label{tab:evaluation}
\end{table*}


\begin{table}[t!]
\centering
\caption{Impact of context length in CA-CLAP retrieval results. The last line randomly chooses 5 utterances before and after the current text as context input of the text encoder.}
\scalebox{.8}{
\begin{tabular}{c|c|c|c|c|c}
\hline \hline 
Context Len & SIM$\uparrow$ & R@1$\uparrow$ & R@5$\uparrow$ & R@10$\uparrow$  & mAP@10$\uparrow$ \\
\hline
0  & 0.517 & 0.574 & 0.822 & 0.890 & 0.680    \\
1  & 0.561 & 0.59  & 0.834 & 0.904 & 0.695   \\
3  & 0.579 & 0.614 & 0.848 & 0.915 & 0.715   \\
5  & \textbf{0.589} & \textbf{0.641} & \textbf{0.865} & \textbf{0.927} & \textbf{0.738} \\
7  & 0.550 & 0.493 & 0.761 & 0.847 & 0.607   \\
10 & 0.499 & 0.345 & 0.622 & 0.736 & 0.464   \\
\hline
random 5 & 0.508 & 0.468 & 0.713 & 0.794 & 0.572  \\
\hline \hline
\end{tabular}
}
\vspace{-0.2cm}
\label{tab:retrieval}
\end{table}

\begin{table}[t!]
\centering
\caption{Effects of speech prompt number for zero-shot TTS. Num denotes the number of speech prompts. Rand denotes random selection and Text refers to the text-only model  MiniLM.}
\scalebox{.7}{
\begin{tabular}{c|p{0.7cm}p{0.7cm}p{0.8cm}|p{0.7cm}p{0.7cm}p{0.7cm}|p{0.7cm}p{0.7cm}p{0.7cm}}
\hline \hline
  & \multicolumn{3}{c|}{\textbf{Energy}$\downarrow$} & \multicolumn{3}{c|}{\textbf{MCD}$\downarrow$} & \multicolumn{3}{c}{\textbf{SECS}$\uparrow$} \\
\hline
Num & Rand & Text & Ours & Rand & Text & Ours & Rand & Text & Ours \\
\hline
1 & 17.817 & 17.612 & 17.333 & 6.655 & 6.583 & 6.507 & 0.726 & 0.730 & 0.734 \\
2 & \textbf{17.375} & \textbf{17.380} & \textbf{16.988} & \textbf{6.507} & \textbf{6.460} & \textbf{6.390} & 0.743 & 0.746 & 0.748 \\
3 & 17.493 & 17.425 & 17.194 & 6.509 & 6.463 & 6.408 & 0.749 & 0.751 & 0.754 \\
4 & 17.571 & 17.762 & 17.460 & 6.546 & 6.516 & 6.462 & \textbf{0.752} & \textbf{0.753} & \textbf{0.756} \\
\hline \hline
\end{tabular}
}
\vspace{-0.5cm}
\label{tab:prompt_num}
\end{table}

\vspace{-0.2cm}
\subsection{Compared Methods}

To evaluate proposed model's performance, we compare the following retrieval methods for prompt-based TTS. 
\begin{itemize}[topsep=2pt, partopsep=0pt, itemsep=0pt, parsep=0pt, leftmargin=*]
  \item \textbf{Self}: use the same text-audio pair in evaluation. This serves as upper bound of prompt-based TTS performance.
  \item \textbf{Random}: randomly select one text-audio pair in the same audiobook as prompt text and prompt speech.
  \item \textbf{Text-only embedding models}: adopt text-only embedding model instead of contrastive multi-modal model to index and query the text-audio pairs. We use the same embedding model in stage indexing and retrieval, and adopt the current text to retrieve.
  For comparison, we adopt the widely used sentence embedding models all-MiniLM-L6-v2\footnote{https://huggingface.co/sentence-transformers/all-MiniLM-L6-v2} and coROM-base\footnote{https://www.modelscope.cn/models/iic/nlp\_corom\_sentence-embedding\_chinese-base}, denoted as \textbf{MiniLM} and \textbf{CoROM}.
\end{itemize}

\vspace{-0.2cm}
\subsection{Objective Evaluation}
To evaluate the effectiveness of our proposed method in terms of naturalness, speaker similarity, and prosodic accuracy, we utilize four metrics including energy, F0, mel-cepstral distortion (MCD), and Speaker Encoder Cosine Similarity (SECS). Noted that we utilize Dynamic Time Warping (DTW) to align generated audio and groundtruth audio before calculation. For assessing speaker similarity, we employ speaker encoder model Resemblyzer\footnote{https://github.com/resemble-ai/Resemblyzer} to compute the SECS between the original and generated speech. The objective evaluation results are presented in Table~\ref{tab:evaluation}.

The results are in line with our expectations: 1) Using the same target speech as the speech prompt yields the best results, as prompt-based TTS can easily replicate the semantic tokens of the prompt. 2) Randomly selecting speech prompts results in the poorest performance across all evaluation metrics, illustrating that choosing a prompt without considering its relevance of desired speaking style can hinder the generation process. 3) Methods using the RAG paradigm can improve generative performance in speaker similarity, prosody, and speaking style compared to non-RAG methods. Moreover, our model, which incorporates context-aware contrastive-based multi-modal embedding, outperforms text-only embedding models (such as MiniLM and CoROM). This suggests that the context-aware style-related embedding generated by the pretrained CA-CLAP model can more precisely match speech prompts with the appropriate speaking style for the desired audio, taking context into account.

\vspace{-0.2cm}
\subsection{Subjective Evaluation}

We conduct two mean opinion score (MOS)  subjective tests including naturalness MOS (NMOS) test to evaluate the naturalness and audio quality, and similarity MOS (SMOS) test to compare the speaking prosody and timbre between groundtruth speech and synthesized speech. We randomly select 30 samples from different audiobooks in the test set. We ask 10 native listeners to rate the NMOS and SMOS on a scale from 1 to 5 with 0.5 point interval. The subjective evaluation results are presented in Table~\ref{tab:evaluation}. The results are consistent with our objective findings, and our proposed method outperforms all the baselines in both NMOS and SMOS. This indicates that the combination of RAG method and context-aware contrastive-based multi-modal embedding extracted from CA-CLAP can match the best speech prompt with suitable speaking style.

\vspace{-0.2cm}
\subsection{Effects of Context Length}

To assess the effectiveness of context information and the impact of context length in the CA-CLAP retrieval process, we adjust context length $l$ from 0 to 10, and evaluate the retrieval results including similarity, recall (R@1 to R@10) and mean average precision (mAP@10) following evaluation in~\cite{clap}. The performance of our CA-CLAP is shown in Table~\ref{tab:retrieval}. We find that the performance first increases from 0 to 5 context length, and decreases when context length is longer than 5. We believe that an appropriate length of context can help with text comprehension, but excessively long context has redundant information which hinders understanding. Moreover, we randomly choose 5 utterances before and after the current text as context and find that it has inferior performance, suggesting that context is indeed helpful in understanding the current text, while incorrect context worsens the performance.

\vspace{-0.1cm}
\subsection{Effects of Speech Prompt Number}

To assess the impact of speech prompt numbers on prompt-based TTS with different RAG methods, we conduct this ablation study, as shown in Table~\ref{tab:prompt_num}. We find that as the prompt number increases, the speaker similarity also gradually increases. This is because the prompt-based TTS tries to clone the speaker timbre and a longer prompt provides more acoustic information. However, the results of energy and MCD show that the performance peaks at a prompt length of 2, and weakens with longer prompts. We believe longer inconsistent content from different prompts may prevent prompt-based TTS from generating coherent speech.

\vspace{-0.3cm}
\section{Conclusion}

In this paper, we present a novel RAG-enhanced prompt-based TTS framework that incorporates a context-aware contrastive language-audio pretraining model. To our knowledge, this is the pioneering zero-shot, prompt-based TTS framework that effectively employs the RAG paradigm and multi-modal context enhancement to refine speech prompt selection and ensure stable generation. To verify the effectiveness of our model, we conduct evaluations on both retrieval and zero-shot TTS. The results indicate that our model outperforms other baselines and can effectively match suitable speech prompts to generate more coherent speech with a context-appropriate speaking style. Additionally, we investigate the impact of context length and the number of speech prompts on our model's performance.

\vspace{-0.1cm}
\section{Acknowledgements}
The work was supported by the National Natural Science Foundation of China (NSFC) (No. 62271083),  the Key Project of the National Language Commission (No. ZDI145-81), and the Fundamental Research Funds for the Central Universities (No. 2023RC73, 2023RC13).


\newpage


\bibliographystyle{IEEEtran}
\bibliography{mybib}

\begin{thebibliography}{10}
\providecommand{\url}[1]{#1}
\csname url@samestyle\endcsname
\providecommand{\newblock}{\relax}
\providecommand{\bibinfo}[2]{#2}
\providecommand{\BIBentrySTDinterwordspacing}{\spaceskip=0pt\relax}
\providecommand{\BIBentryALTinterwordstretchfactor}{4}
\providecommand{\BIBentryALTinterwordspacing}{\spaceskip=\fontdimen2\font plus
\BIBentryALTinterwordstretchfactor\fontdimen3\font minus \fontdimen4\font\relax}
\providecommand{\BIBforeignlanguage}[2]{{%
\expandafter\ifx\csname l@#1\endcsname\relax
\typeout{** WARNING: IEEEtran.bst: No hyphenation pattern has been}%
\typeout{** loaded for the language `#1'. Using the pattern for}%
\typeout{** the default language instead.}%
\else
\language=\csname l@#1\endcsname
\fi
#2}}
\providecommand{\BIBdecl}{\relax}
\BIBdecl

\bibitem{soundstream}
\BIBentryALTinterwordspacing
N.~Zeghidour, A.~Luebs, A.~Omran, J.~Skoglund, and M.~Tagliasacchi, ``Soundstream: An end-to-end neural audio codec,'' \emph{{IEEE} {ACM} Trans. Audio Speech Lang. Process.}, vol.~30, pp. 495--507, 2022. [Online]. Available: \url{https://doi.org/10.1109/TASLP.2021.3129994}
\BIBentrySTDinterwordspacing

\bibitem{encodec}
A.~D{\'e}fossez, J.~Copet, G.~Synnaeve, and Y.~Adi, ``High fidelity neural audio compression,'' \emph{arXiv preprint arXiv:2210.13438}, 2022.

\bibitem{hubert}
\BIBentryALTinterwordspacing
W.~Hsu, B.~Bolte, Y.~H. Tsai, K.~Lakhotia, R.~Salakhutdinov, and A.~Mohamed, ``Hubert: Self-supervised speech representation learning by masked prediction of hidden units,'' \emph{{IEEE} {ACM} Trans. Audio Speech Lang. Process.}, vol.~29, pp. 3451--3460, 2021. [Online]. Available: \url{https://doi.org/10.1109/TASLP.2021.3122291}
\BIBentrySTDinterwordspacing

\bibitem{megatts}
\BIBentryALTinterwordspacing
Z.~Jiang, Y.~Ren, Z.~Ye, J.~Liu, C.~Zhang, Q.~Yang, S.~Ji, R.~Huang, C.~Wang, X.~Yin, Z.~Ma, and Z.~Zhao, ``Mega-tts: Zero-shot text-to-speech at scale with intrinsic inductive bias,'' \emph{CoRR}, vol. abs/2306.03509, 2023. [Online]. Available: \url{https://doi.org/10.48550/arXiv.2306.03509}
\BIBentrySTDinterwordspacing

\bibitem{megatts2}
\BIBentryALTinterwordspacing
Z.~Jiang, J.~Liu, Y.~Ren, J.~He, C.~Zhang, Z.~Ye, P.~Wei, C.~Wang, X.~Yin, Z.~Ma, and Z.~Zhao, ``Mega-tts 2: Zero-shot text-to-speech with arbitrary length speech prompts,'' \emph{CoRR}, vol. abs/2307.07218, 2023. [Online]. Available: \url{https://doi.org/10.48550/arXiv.2307.07218}
\BIBentrySTDinterwordspacing

\bibitem{valle}
C.~Wang, S.~Chen, Y.~Wu, Z.~Zhang, L.~Zhou, S.~Liu, Z.~Chen, Y.~Liu, H.~Wang, J.~Li \emph{et~al.}, ``Neural codec language models are zero-shot text to speech synthesizers,'' \emph{arXiv preprint arXiv:2301.02111}, 2023.

\bibitem{audiolm}
Z.~Borsos, R.~Marinier, D.~Vincent, E.~Kharitonov, O.~Pietquin, M.~Sharifi, D.~Roblek, O.~Teboul, D.~Grangier, M.~Tagliasacchi \emph{et~al.}, ``Audiolm: a language modeling approach to audio generation,'' \emph{IEEE/ACM Transactions on Audio, Speech, and Language Processing}, 2023.

\bibitem{naturespeech2}
\BIBentryALTinterwordspacing
K.~Shen, Z.~Ju, X.~Tan, Y.~Liu, Y.~Leng, L.~He, T.~Qin, S.~Zhao, and J.~Bian, ``Naturalspeech 2: Latent diffusion models are natural and zero-shot speech and singing synthesizers,'' \emph{CoRR}, vol. abs/2304.09116, 2023. [Online]. Available: \url{https://doi.org/10.48550/arXiv.2304.09116}
\BIBentrySTDinterwordspacing

\bibitem{speartts}
E.~Kharitonov, D.~Vincent, Z.~Borsos, R.~Marinier, S.~Girgin, O.~Pietquin, M.~Sharifi, M.~Tagliasacchi, and N.~Zeghidour, ``Speak, read and prompt: High-fidelity text-to-speech with minimal supervision,'' \emph{arXiv preprint arXiv:2302.03540}, 2023.

\bibitem{w2vbert}
Y.-A. Chung, Y.~Zhang, W.~Han, C.-C. Chiu, J.~Qin, R.~Pang, and Y.~Wu, ``W2v-bert: Combining contrastive learning and masked language modeling for self-supervised speech pre-training,'' in \emph{2021 IEEE Automatic Speech Recognition and Understanding Workshop (ASRU)}.\hskip 1em plus 0.5em minus 0.4em\relax IEEE, 2021, pp. 244--250.

\bibitem{fastspeech2}
Y.~Ren, C.~Hu, X.~Tan, T.~Qin, S.~Zhao, Z.~Zhao, and T.-Y. Liu, ``Fastspeech 2: Fast and high-quality end-to-end text to speech,'' \emph{arXiv preprint arXiv:2006.04558}, 2020.

\bibitem{tacotron}
Y.~Wang, R.~Skerry-Ryan, D.~Stanton, Y.~Wu, R.~J. Weiss, N.~Jaitly, Z.~Yang, Y.~Xiao, Z.~Chen, S.~Bengio \emph{et~al.}, ``Tacotron: Towards end-to-end speech synthesis,'' \emph{arXiv preprint arXiv:1703.10135}, 2017.

\bibitem{cot}
J.~Wei, X.~Wang, D.~Schuurmans, M.~Bosma, F.~Xia, E.~Chi, Q.~V. Le, D.~Zhou \emph{et~al.}, ``Chain-of-thought prompting elicits reasoning in large language models,'' \emph{Advances in Neural Information Processing Systems}, vol.~35, pp. 24\,824--24\,837, 2022.

\bibitem{zerocot}
T.~Kojima, S.~S. Gu, M.~Reid, Y.~Matsuo, and Y.~Iwasawa, ``Large language models are zero-shot reasoners,'' \emph{Advances in neural information processing systems}, vol.~35, pp. 22\,199--22\,213, 2022.

\bibitem{xue2022paratts}
L.~Xue, F.~K. Soong, S.~Zhang, and L.~Xie, ``Paratts: Learning linguistic and prosodic cross-sentence information in paragraph-based tts,'' \emph{IEEE/ACM Transactions on Audio, Speech, and Language Processing}, vol.~30, pp. 2854--2864, 2022.

\bibitem{chen2022unsupervised}
X.~Chen, S.~Lei, Z.~Wu, D.~Xu, W.~Zhao, and H.~Meng, ``Unsupervised multi-scale expressive speaking style modeling with hierarchical context information for audiobook speech synthesis,'' in \emph{Proceedings of the 29th International Conference on Computational Linguistics}, 2022, pp. 7193--7202.

\bibitem{chen2023stylespeech}
X.~Chen, X.~Wang, S.~Zhang, L.~He, Z.~Wu, X.~Wu, and H.~Meng, ``Stylespeech: Self-supervised style enhancing with vq-vae-based pre-training for expressive audiobook speech synthesis,'' \emph{arXiv preprint arXiv:2312.12181}, 2023.

\bibitem{m2ctts}
J.~Xue, Y.~Deng, F.~Wang, Y.~Li, Y.~Gao, J.~Tao, J.~Sun, and J.~Liang, ``M2-ctts: End-to-end multi-scale multi-modal conversational text-to-speech synthesis,'' in \emph{ICASSP 2023-2023 IEEE International Conference on Acoustics, Speech and Signal Processing (ICASSP)}.\hskip 1em plus 0.5em minus 0.4em\relax IEEE, 2023, pp. 1--5.

\bibitem{deng2023cmcu}
Y.~Deng, J.~Xue, F.~Wang, Y.~Gao, and Y.~Li, ``Cmcu-css: Enhancing naturalness via commonsense-based multi-modal context understanding in conversational speech synthesis,'' in \emph{Proceedings of the 31st ACM International Conference on Multimedia}, 2023, pp. 6081--6089.

\bibitem{deng2023concss}
Y.~Deng, J.~Xue, Y.~Jia, Q.~Li, Y.~Han, F.~Wang, Y.~Gao, D.~Ke, and Y.~Li, ``Concss: Contrastive-based context comprehension for dialogue-appropriate prosody in conversational speech synthesis,'' \emph{arXiv preprint arXiv:2312.10358}, 2023.

\bibitem{rag4ki}
\BIBentryALTinterwordspacing
P.~S.~H. Lewis, E.~Perez, A.~Piktus, F.~Petroni, V.~Karpukhin, N.~Goyal, H.~K{\"{u}}ttler, M.~Lewis, W.~Yih, T.~Rockt{\"{a}}schel, S.~Riedel, and D.~Kiela, ``Retrieval-augmented generation for knowledge-intensive {NLP} tasks,'' in \emph{Advances in Neural Information Processing Systems 33: Annual Conference on Neural Information Processing Systems 2020, NeurIPS 2020, December 6-12, 2020, virtual}, H.~Larochelle, M.~Ranzato, R.~Hadsell, M.~Balcan, and H.~Lin, Eds., 2020. [Online]. Available: \url{https://proceedings.neurips.cc/paper/2020/hash/6b493230205f780e1bc26945df7481e5-Abstract.html}
\BIBentrySTDinterwordspacing

\bibitem{ragmp}
\BIBentryALTinterwordspacing
K.~Guu, K.~Lee, Z.~Tung, P.~Pasupat, and M.~Chang, ``Retrieval augmented language model pre-training,'' in \emph{Proceedings of the 37th International Conference on Machine Learning, {ICML} 2020, 13-18 July 2020, Virtual Event}, ser. Proceedings of Machine Learning Research, vol. 119.\hskip 1em plus 0.5em minus 0.4em\relax {PMLR}, 2020, pp. 3929--3938. [Online]. Available: \url{http://proceedings.mlr.press/v119/guu20a.html}
\BIBentrySTDinterwordspacing

\bibitem{clap}
\BIBentryALTinterwordspacing
Y.~Wu, K.~Chen, T.~Zhang, Y.~Hui, T.~Berg{-}Kirkpatrick, and S.~Dubnov, ``Large-scale contrastive language-audio pretraining with feature fusion and keyword-to-caption augmentation,'' in \emph{{IEEE} International Conference on Acoustics, Speech and Signal Processing {ICASSP} 2023, Rhodes Island, Greece, June 4-10, 2023}.\hskip 1em plus 0.5em minus 0.4em\relax {IEEE}, 2023, pp. 1--5. [Online]. Available: \url{https://doi.org/10.1109/ICASSP49357.2023.10095969}
\BIBentrySTDinterwordspacing

\bibitem{vits}
\BIBentryALTinterwordspacing
J.~Kim, J.~Kong, and J.~Son, ``Conditional variational autoencoder with adversarial learning for end-to-end text-to-speech,'' in \emph{Proceedings of the 38th International Conference on Machine Learning, {ICML} 2021, 18-24 July 2021, Virtual Event}, ser. Proceedings of Machine Learning Research, M.~Meila and T.~Zhang, Eds., vol. 139.\hskip 1em plus 0.5em minus 0.4em\relax {PMLR}, 2021, pp. 5530--5540. [Online]. Available: \url{http://proceedings.mlr.press/v139/kim21f.html}
\BIBentrySTDinterwordspacing

\bibitem{transferTTS}
\BIBentryALTinterwordspacing
M.~Kim, M.~Jeong, B.~J. Choi, S.~Ahn, J.~Y. Lee, and N.~S. Kim, ``Transfer learning framework for low-resource text-to-speech using a large-scale unlabeled speech corpus,'' in \emph{Interspeech 2022, 23rd Annual Conference of the International Speech Communication Association, Incheon, Korea, 18-22 September 2022}, H.~Ko and J.~H.~L. Hansen, Eds.\hskip 1em plus 0.5em minus 0.4em\relax {ISCA}, 2022, pp. 788--792. [Online]. Available: \url{https://doi.org/10.21437/Interspeech.2022-225}
\BIBentrySTDinterwordspacing

\bibitem{chen2022htsat}
K.~Chen, X.~Du, B.~Zhu, Z.~Ma, T.~Berg-Kirkpatrick, and S.~Dubnov, ``Hts-at: A hierarchical token-semantic audio transformer for sound classification and detection,'' in \emph{ICASSP 2022-2022 IEEE International Conference on Acoustics, Speech and Signal Processing (ICASSP)}.\hskip 1em plus 0.5em minus 0.4em\relax IEEE, 2022, pp. 646--650.

\end{thebibliography}

\end{document}